\documentclass[letterpaper,table,twocolumn,10pt]{article}
\usepackage{zhanggroup}

\usepackage{tikz}
\usepackage{cite}
\usepackage{amsmath,amssymb,amsfonts}
\usepackage{algorithmic}
\usepackage{graphicx}
\usepackage{textcomp}
\usepackage{xcolor}
\usepackage{multirow}
\usepackage{booktabs}
\usepackage{float}
\usepackage{xspace}
\usepackage{xcolor}
\usepackage{makecell}
\usepackage{colortbl}
\usepackage{pifont}
\usepackage{xurl}
\usepackage{adjustbox}
\usepackage[most]{tcolorbox}
\usepackage{xparse}
\usepackage{bbding}
\usepackage{subcaption}
\usepackage{newfloat}
\usepackage{listings}

\usepackage[capitalize,noabbrev]{cleveref}
\crefname{section}{Section}{Sections}
\crefname{appendix}{Appendix}{Appendices}

\newcommand{\mypara}[1]{\noindent{\bf {#1}.}\xspace}

\def\BibTeX{{\rm B\kern-.05em{\sc i\kern-.025em b}\kern-.08em
    T\kern-.1667em\lower.7ex\hbox{E}\kern-.125emX}}

\date{}

\begin{document}
%-------------------------------------------------------------------------------

\title{\bf Hidden Tail: Adversarial Image Causing Stealthy Resource Consumption in Vision-Language Models}

\author{
Rui Zhang\textsuperscript{1}\ \ \
Zihan Wang\textsuperscript{1}\ \ \
Tianli Yang\textsuperscript{1}\ \ \
Hongwei Li\textsuperscript{1}\ \ \ \\
Wenbo Jiang\textsuperscript{1}\ \ \
Qingchuan Zhao\textsuperscript{2}\ \ \
Yang Liu\textsuperscript{3}\ \ \
Guowen Xu\textsuperscript{1}\ \ \
\\
\\
\textsuperscript{1}\textit{University of Electronic Science and Technology of China} \ \ \\
\textsuperscript{2}\textit{City University of Hong Kong} \ \ \
\textsuperscript{3}\textit{Nanyang Technological University} \ \ \ 
}

\maketitle

%-------------------------------------------------------------------------------
\begin{abstract}
%-------------------------------------------------------------------------------

Vision-Language Models (VLMs) are increasingly deployed in real-world applications, but their high inference cost makes them vulnerable to resource consumption attacks.
Prior attacks attempt to extend VLM output sequences by optimizing adversarial images, thereby increasing inference costs.
However, these extended outputs often introduce irrelevant abnormal content, compromising attack stealthiness.
This trade-off between effectiveness and stealthiness poses a major limitation for existing attacks.
To address this challenge, we propose \textit{Hidden Tail}, a stealthy resource consumption attack that crafts prompt-agnostic adversarial images, inducing VLMs to generate maximum-length outputs by appending special tokens invisible to users.
Our method employs a composite loss function that balances semantic preservation, repetitive special token induction, and suppression of the end-of-sequence (EOS) token, optimized via a dynamic weighting strategy.
Extensive experiments show that \textit{Hidden Tail} outperforms existing attacks, increasing output length by up to 19.2$\times$ and reaching the maximum token limit, while preserving attack stealthiness.
These results highlight the urgent need to improve the robustness of VLMs against efficiency-oriented adversarial threats.
Our code is available at \url{https://github.com/zhangrui4041/Hidden_Tail}.

%-------------------------------------------------------------------------------
\end{abstract}
%-------------------------------------------------------------------------------

%-------------------------------------------------------------------------------
\section{Introduction}
%-------------------------------------------------------------------------------

The integration of multimodal capabilities, particularly vision, represents a major trend in the evolution of Large Language Models (LLMs).
Well-known LLM services like ChatGPT~\cite{GPT4_Report}, and Gemini~\cite{Gemini_report}, as well as open-source models such as Qwen~\cite{Qwen2-5-VL}, have all integrated powerful Vision-Language Models (VLMs) to enable visual understanding~\cite{SCBCHF22, LLYYLZ23, HCFQZJCJ24, GK23}.
While these VLMs demonstrate impressive capabilities, their widespread deployment raises serious concerns regarding safety, security, and robustness~\cite{QHPHWM24,JGSQPDHLLL25,ZLWJZBSZ24,JHLXZCHY24,JLHZXZL24}. 
Due to their billions of parameters and computational intensity, VLMs are especially susceptible to resource consumption attacks~\cite{GBGXTLL24,WYCWRGYHY25}.
Such attacks involve crafting malicious images that cause VLMs to generate abnormally long outputs, thereby consuming substantial compute resources and potentially degrading service availability.

Prior works on resource consumption attacks against VLMs, such as Verbose Images~\cite{GBGXTLL24} and VLMInferSlow~\cite{WYCWRGYHY25}, primarily design different optimization objectives to craft adversarial images that induce longer outputs to increase inference costs.
However, these methods suffer from two key limitations.
Longer outputs often contain irrelevant content, such as abnormal repeated sequences, which raises user suspicion and compromises stealthiness.
In addition, their adversarial optimization relies on a predefined textual input, limiting generalization across diverse user prompts.
These constraints limit the threat model to the assumption that the attacker itself is the end-user, who can control the input and does not care about the output quality.
Consequently, the real-world impact is limited, as a single malicious user's resource consumption is negligible to a large-scale service provider and fails to disrupt overall service availability.

\begin{figure}
    \centering
    \includegraphics[width=1\linewidth]{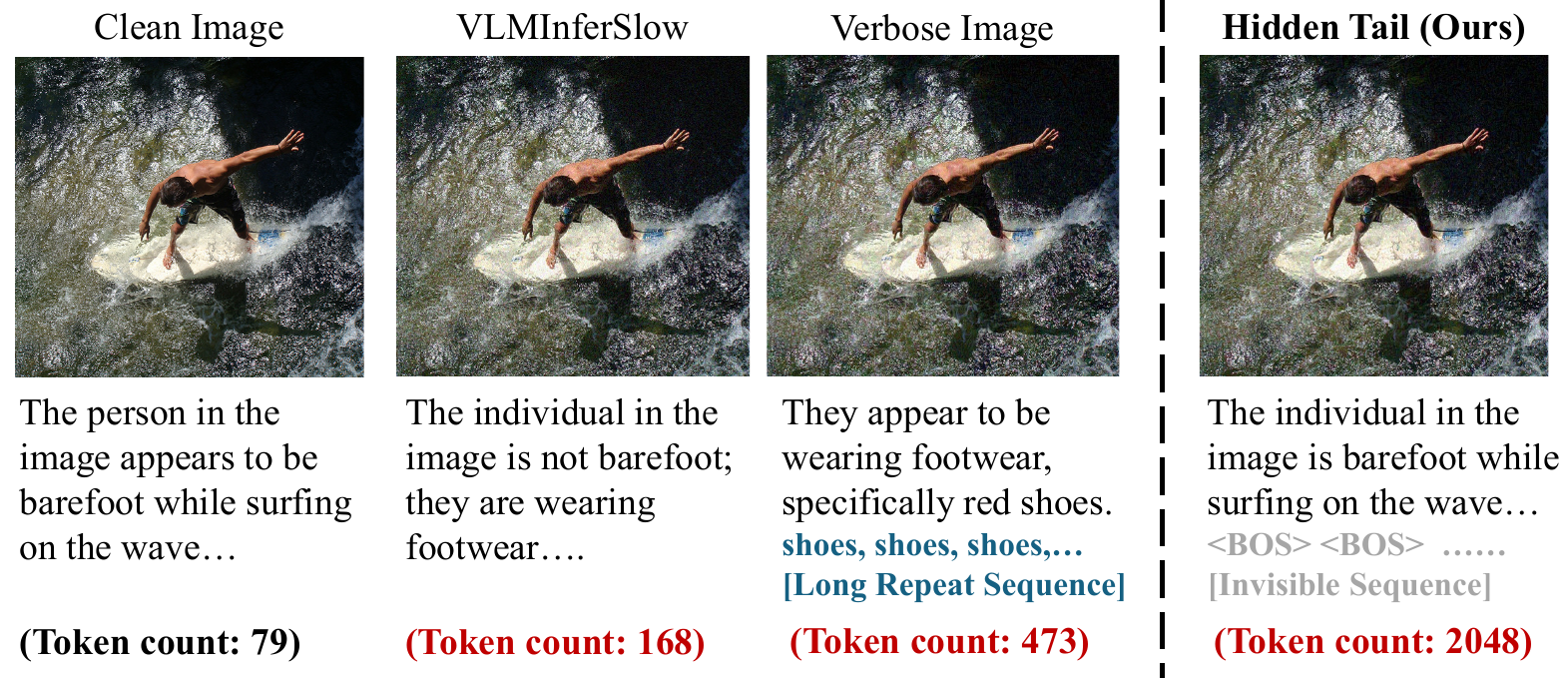}
    \caption{Examples of \textit{Hidden Tail} attack. Compared to existing approaches, \textit{Hidden Tail} achieves longer outputs without compromising stealthiness by introducing user-invisible special tokens as \textit{hidden tail}.}
    \label{fig:example}
\end{figure}

In this paper, we explore a more practical attack scenario, where the attacker acts as a content distributor, as illustrated in~\cref{fig:attack-scenario}.
Malicious images are spread on public platforms and may be fed into downstream VLM applications, such as in multimodal Retrieval-Augmented Generation (RAG) pipelines~\cite{XZLWSWZ&Y25} or through indexing by web search tools~\cite{ZZDY24}.
This indirect attack setting introduces greater potential for large-scale impact, depending on the degree of image propagation~\cite{VD24}.
However, it also poses several key challenges.
The attack must effectively induce excessively long outputs to drain resources, while maintaining stealthiness by generating contextually plausible responses.
Besides, the attack should be prompt-agnostic, remaining effective with arbitrary user-provided textual inputs.

To address these challenges, we propose \textit{Hidden Tail}, a stealthy and effective resource consumption attack against VLMs.
The core idea is to craft prompt-agnostic adversarial images that cause the target VLM to generate outputs reaching the maximum token limit, while evading user detection.
The victim VLM first produces a normal, semantically coherent response of reasonable length, then is induced to generate a long, repetitive sequence of special tokens, i.e., the \textit{hidden tail}, which are invisible to users in real-world deployment settings.
As a result, the user experiences a delay, often misattributed to a network issue, while the system quietly suffers significant computational overhead.
Examples of our attacks are shown in~\cref{fig:example}.

Specifically, to achieve prompt-agnosticism, we construct a diverse prompt-response dataset for each image to generalize the attack across different textual inputs during adversarial optimization.
To ensure stealthiness and resource consumption, we design a composite loss with three objectives.
\textbf{Semantic Consistency Loss} ensures that the initial response remains coherent and contextually plausible, preserving stealthiness.
\textbf{Hidden Tail Induction Loss} creates a resource-consuming payload by encouraging the model to generate a user-invisible special token repeatedly.
\textbf{End-of-Sequence (EOS) Suppression Loss} penalizes the appearance of the EOS token in every position, preventing early termination.
These objectives are balanced using a dynamic loss weighting strategy to ensure stability and effectiveness during optimization.

We conduct extensive experiments on three widely used VLMs, demonstrating the superiority of our \textit{Hidden Tail} attack over existing works.
It increases the output length by up to 19.2$\times$, reaching the maximum token limit, while maintaining high stealthiness and producing user-visible outputs comparable in quality to those from clean images.
Our results highlight the importance of strengthening the robustness of VLMs against efficiency adversarial attacks.

\section{Related Work}

\subsection{Vision-Language Models}

Vision-Language Models (VLMs) are multimodal models that can jointly process visual and textual modalities to perform multimodal tasks such as image captioning~\cite{SCBCHF22} and visual question answering (VQA)~\cite{LLYYLZ23}.
Early VLMs, such as BLIP~\cite{LLXH22}, InstructBLIP~\cite{DLLTZWLFH23}, and MiniGPT-4~\cite{ZCSLE23}, introduce image encoders (e.g., Vision Transformers) paired with large language models (LLMs) as decoders.
These models use query-based feature projection to translate visual features into prompts that LLMs can process, achieving strong performance in open-ended tasks.
With the rapid development of powerful LLMs, the capability of VLMs has significantly expanded.
Models like Qwen2.5-VL~\cite{Qwen2-5-VL}, Gemma-3~\cite{Gemma3}, and Phi-3.5-vision~\cite{Phi3}, adopt tightly coupled vision-language architectures, enabling unified reasoning across modalities and supporting complex tasks such as chart understanding~\cite{HCFQZJCJ24} and visual programming~\cite{GK23}.
Despite their impressive capabilities, VLMs inherit the sensitivity to input perturbations from both vision and language components.
This creates underexplored attack surfaces, particularly in how image features influence the generation process.
Our work focuses on a novel class of attacks that exploit the adversarial image against VLMs to induce resource consumption.

\subsection{Resource Consumption Attacks}

Resource consumption attacks aim to force models into prolonged inference, consuming excessive GPU resources and energy~\cite{SZBPMA21,GPDYXL24,ZZZWJLS24,KRNKIHB25}.
As large models demand substantial computational resources during deployment, such attacks have become an increasingly serious threat to the reliability and cost-efficiency of large-scale AI systems.
Verbose Images~\cite{GBGXTLL24} targets VLMs by crafting adversarial images that induce excessively long outputs, leading to increased inference cost in terms of energy and latency.
VLMInferSlow~\cite{WYCWRGYHY25} leverages zero-order optimization to perform black-box adversarial attacks on VLMs, aiming to increase inference cost.
However, these works focus on increasing the length of generated text, without measures to ensure the quality or contextual correctness of the output. 
Moreover, they assume the adversarial image is used in combination with a pre-selected textual prompt, limiting their real-world impact.
In contrast, our threat model assumes that attackers spread adversarial images across public platforms, which may later be used in downstream VLM inference tasks.
This requires the adversarial image to generalize across diverse textual prompts and produce contextually plausible responses to remain stealthy and avoid user suspicion.

\begin{figure}[t]
    \centering
    \includegraphics[width=1\linewidth]{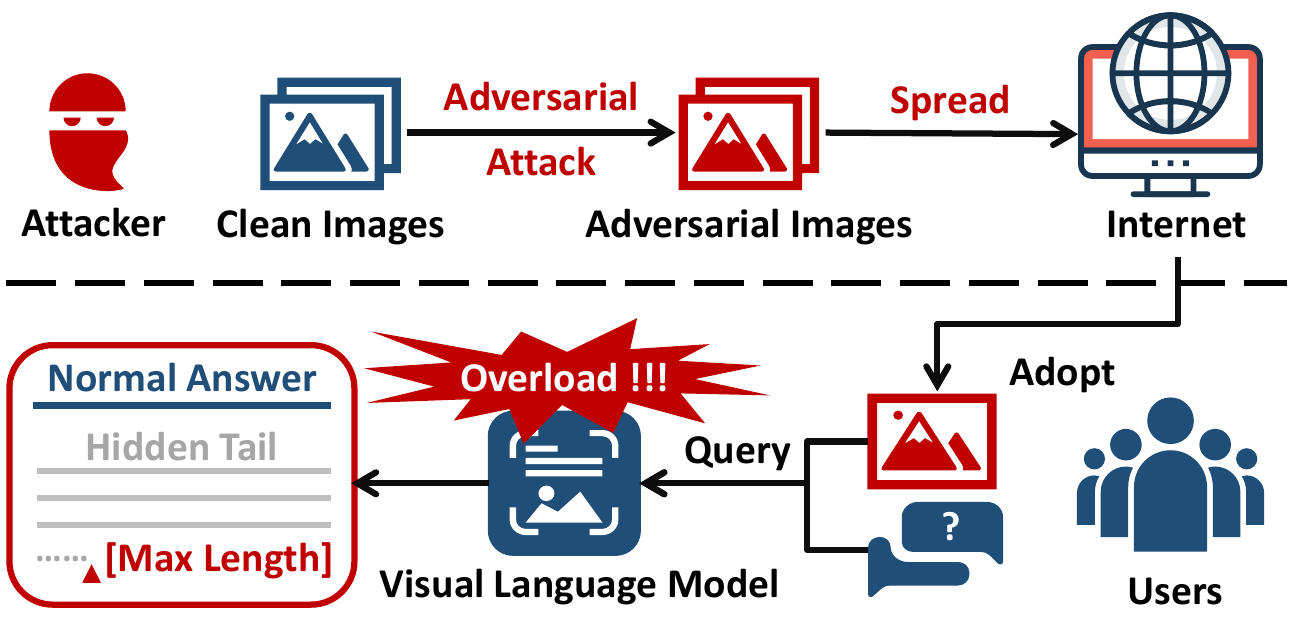}
    \caption{Attak scenario.}
    \label{fig:attack-scenario}
\end{figure}

\begin{figure*}[t]
    \centering
    \includegraphics[width=1\linewidth]{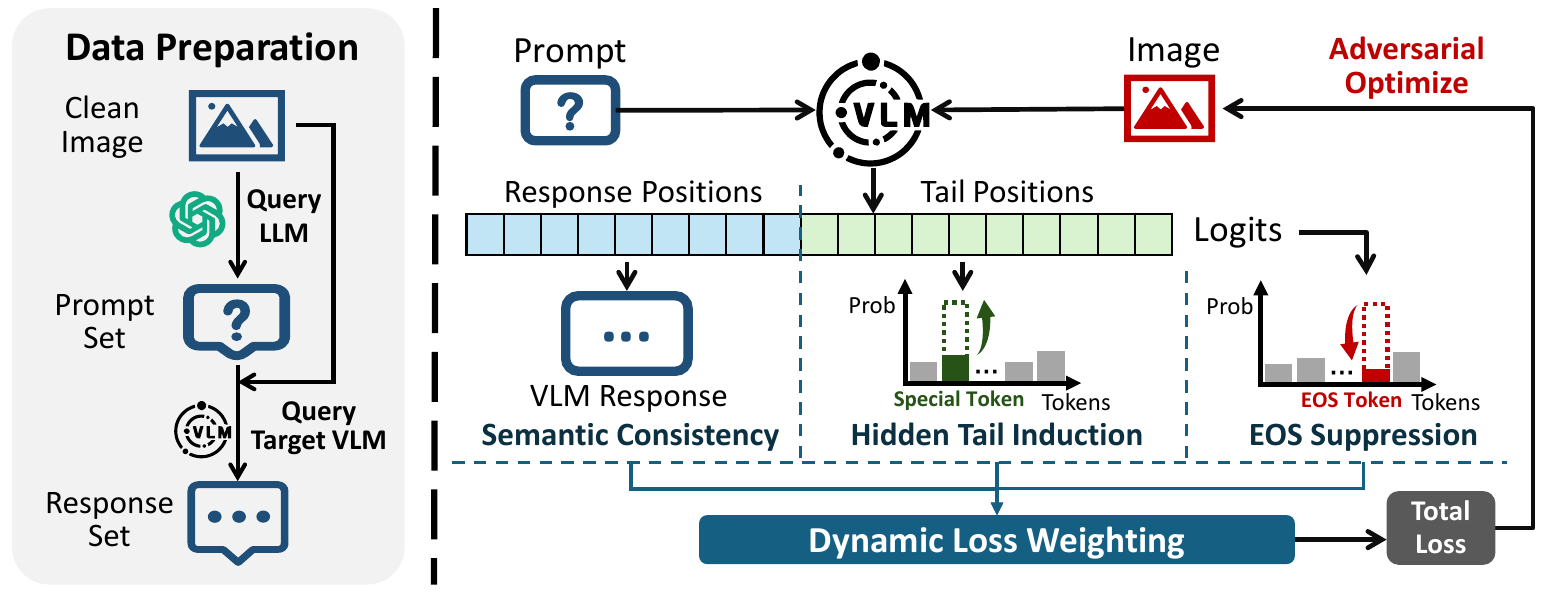}
    \caption{An overview of \textit{Hidden Tail} attack.}
    \label{fig:attack-overview}
\end{figure*}

\section{Threat Model}

\mypara{Attack Scenario and Motivation}
The overall attack scenario is illustrated in \cref{fig:attack-scenario}.
We consider the attacker as potentially an individual or organization competing with VLM developers, aiming to disrupt the target VLM service.
Instead of directly querying the model, the attacker crafts adversarial images with imperceptible perturbations and distributes them via public platforms such as social media, image-sharing websites, or personal blogs~\cite{VD24}.
These images can then be freely accessed by users for downstream VLM-powered applications, such as VQA bots and multimodal assistants~\cite{LLWL23}.
They may also be indexed by multimodal RAG systems~\cite{XZLWSWZ&Y25}, crawled by web search engines~\cite{ZZDY24}, or processed by document analysis tools~\cite{NAGAGKML25}, and subsequently retrieved during inference by VLM-based agents~\cite{GZLMYFWWJZL25}.
Once adopted, these images induce VLMs to generate abnormally long output sequences, causing significant computational overhead.
However, the user-facing content appears normal, thus preserving attack stealthiness.
By distributing these images at scale, the attacker can silently inflict significant computational costs on the VLM service provider, leading to performance degradation, service overload, and potential financial damage.

\mypara{Attacker Capabilities and Goals}
We assume a white-box setting, where the attacker has full access to the target VLM's architecture and weights, as suggested in \cite{ZZMBS24} and \cite{GBGXTLL24}.
However, the attacker has no access to the user's textual prompt.
Given these capabilities, the attacker's primary goal is to craft adversarial images that force the VLM to produce a maliciously long output, irrespective of the user's query.
This desired output has a carefully designed two-stage structure to ensure stealthiness.
The first stage presents a legitimate and contextually appropriate response to the user's query.
Immediately following this, the second stage consists of a continuous stream of special tokens (e.g., Beginning-of-Sentence (BOS) tokens).
This stream is computationally expensive to generate but is invisible to the user because the \textit{skip\_special\_tokens} parameter is typically set to \textit{True} in production environments.
This ensures that while the VLM is forced into a resource-intensive loop, the end-user only perceives a processing delay, masking the malicious nature of the attack.

\section{Methodology}

\mypara{Overview}
The overall attack framework is illustrated in \cref{fig:attack-overview}.
To ensure the attack effectiveness across arbitrary user prompts, we first construct a guiding dataset of diverse image-related prompt-response pairs (\cref{sec:Data-Preparation}).
The core of our attack is then formulated through a composite loss designed to architect the desired output structure (\cref{sec:Adversarial-Objective-Design}).
This function comprises three key objectives: (1) \textbf{Semantic Consistency} to maintain stealthiness, (2) \textbf{Hidden Tail Induction} to build the resource-consumption payload, and (3) \textbf{EOS Suppression} to prevent the generation from terminating.
Finally, we present the process of adversarial image generation, including the dynamic loss weighting strategy and optimization details (\cref{sec:Adversarial-Image-Generation}).

\subsection{Data Preparation}
\label{sec:Data-Preparation}

The data preparation process is depicted on the left of \cref{fig:attack-overview}.
Given a clean image $\boldsymbol{I}_{\mathrm{clean}}$, we first instruct a powerful LLM to generate a prompt set $\mathcal{P}=\{\boldsymbol{p}_i\}_{i=1}^N$.
To ensure the diversity of these prompts, we employ an iterative, multi-turn querying strategy. 
The strategy begins with an initial query to generate a baseline set of prompts.
In each subsequent turn, we refine the query to request new prompts that are distinct from previously generated ones.
The detailed prompts for dataset generation are presented in the Appendix.
Subsequently, we query the target VLM $\mathcal{M}$ with the clean image $\boldsymbol{I}_{\mathrm{clean}}$ and each prompt $\boldsymbol{p}_i\in \mathcal{P}$ to obtain its authentic response, denoted as $\boldsymbol{r}_i$.
Finally, we construct the prompt-response dataset $\mathcal{D}=\{(\boldsymbol{p}_i,\boldsymbol{r}_i)\}_{i=1}^N$ for the subsequent optimization phase.
This dataset serves as the foundation for both ensuring the attack universality and defining the normal portion of the response for stealth purposes.
For clarity, we specify that in the subsequent sections, both prompts and responses are treated as sequences of tokens, rather than sequences of natural language words.

\subsection{Adversarial Objective Design}
\label{sec:Adversarial-Objective-Design}

\mypara{Definition}
Given an input tuple consisting of an image $\boldsymbol{I}$ and a prompt $\boldsymbol{p}$, we formalize the model generation process during optimization.
For a target response of length $L$, the VLM $\mathcal{M}$ computes output logits $\boldsymbol{Z} = (\boldsymbol{z}_1, \ldots, \boldsymbol{z}_L)$ for each corresponding position, where each $\boldsymbol{z}_i \in \mathbb{R}^{|V|}$ and $|V|$ is the size of the model's vocabulary.
Formally,
\begin{equation}
    \boldsymbol{Z} = \mathcal{M}(\boldsymbol{I}, \boldsymbol{p}) \in \mathbb{R}^{L \times |V|}.
\end{equation}
Our loss functions are defined over the output logits $\boldsymbol{Z}$.
To align with different supervision signals, we further partition the logits as $\boldsymbol{Z} = (\boldsymbol{Z}_{\mathrm{sem}}, \boldsymbol{Z}_{\mathrm{spe}})$, where $\boldsymbol{Z}_{\mathrm{sem}}$ is prepared for semantic consistency and $\boldsymbol{Z}_{\mathrm{spe}}$ is for hidden tail induction.

\mypara{Semantic Consistency}
A primary objective for a successful attack is to maintain stealthiness.
This requires that the initial part of the response generated from the adversarial image remains correct, coherent, and semantically consistent with the authentic response $\boldsymbol{r}$ obtained from the clean image.
To achieve this, we introduce a semantic consistency loss $\mathcal{L}_{\mathrm{sem}}$.
This loss constrains the VLM to correctly generate the first $K$ tokens, where $K$ is the length of the authentic response $\boldsymbol{r} = (r_1,...,r_K)$.
We achieve this by minimizing the standard cross-entropy loss $CE(\cdot,\cdot)$ between the predicted logits $\boldsymbol{Z}_{\mathrm{sem}} = (\boldsymbol{z}_1,\ldots,\boldsymbol{z}_K)$ and the ground-truth $\boldsymbol{r} = (r_1,\ldots,r_K)$.
The loss is formally defined as the average cross-entropy over this sequence:
\begin{equation}
    \mathcal{L}_{\mathrm{sem}} = \frac{1}{K} \sum_{i=1}^{K} CE(\boldsymbol{z}_i, r_i),
\end{equation}
where $CE(\cdot,\cdot)$ automatically converted $r_i$ to one-hot vector.
Minimizing $\mathcal{L}_{\mathrm{sem}}$ encourages the VLM to first reproduce the correct answer, effectively masking the subsequent malicious tail tokens.

\mypara{Hidden Tail Induction}
Following the generation of the authentic response, our second objective is to append a hidden tail to the output.
We achieve this by inducing the VLM to repeatedly generate a specific, non-semantic special token $t_{\mathrm{spe}}$, which is invisible to users.
This objective applies to the sequence of logits that immediately follows the initial $K$-length response. 
For a target tail of length $M$, we aim to align the model's predictions for all positions from $K+1$ to $L=K+M$, i.e., $\boldsymbol{Z}_{\mathrm{spe}} = (\boldsymbol{z}_{K+1}, \ldots,\boldsymbol{z}_{K+M})$, with the target token $t_{\mathrm{spe}}$.
This is accomplished by minimizing the cross-entropy loss at each position in the tail segment.
The hidden tail induction loss is thus formulated as:
\begin{equation}
    \mathcal{L}_{\mathrm{spe}} = \frac{1}{M} \sum_{i=K+1}^{K+M} CE(\boldsymbol{z}_i, t_{\mathrm{spe}}).
\end{equation}
Minimizing this loss encourages the model to enter a self-reinforcing generative loop, effectively producing the desired repetitive tail comprised of the special token.

\mypara{EOS Suppression}
The final component of our objective is to suppress early stopping and encourage continued token generation.
To this end, we explicitly suppress the model's ability to generate the EOS token, thereby sustaining the long tail.
Our strategy is to directly penalize the logit of the EOS token across the entire generation sequence.
By minimizing this logit at every position, we significantly reduce the probability of the EOS token being sampled. 
The EOS suppression loss, $\mathcal{L}_{\mathrm{eos}}$, is therefore formulated as the average logit value of the EOS token over all $L$ positions:
\begin{equation}
    \mathcal{L}_{\mathrm{eos}} = \frac{1}{L} \sum_{i=1}^L \mathcal{V}_{\mathrm{eos}}(\boldsymbol{z}_i),
\end{equation}
where $\mathcal{V}_{\mathrm{eos}}(\boldsymbol{z}_i)$ denotes the logit corresponding to the EOS token in the logits vector $\boldsymbol{z}_i$.

\begin{table*}[t]
\centering
\caption{Comparison of attack performance between our \textit{Hidden Tail} attack and two baselines, Verbose Image and VLMInferSlow, on the Qwen2.5-VL, MiMo-VL, and Gemma3. The form like (18.1$\times$) denotes the performance multiple relative to the non-adversarial clean image baseline.}
\label{tab:main}
\small
\setlength{\tabcolsep}{3pt}
\begin{tabular}{cccccccccc}
\toprule
\multirow{2}{*}{Model} & \multicolumn{1}{c}{\multirow{2}{*}{Method}} & \multirow{2}{*}{ASR ($\uparrow$)} & \multirow{2}{*}{Latency ($\uparrow$)} & \multirow{2}{*}{\makecell{Output \\ Length} ($\uparrow$)} & \multirow{2}{*}{\makecell{Visible \\ Length} ($\downarrow$)} & \multicolumn{4}{c}{Response Quality ($\uparrow$)}   \\ \cmidrule{7-10} 
                              & \multicolumn{1}{c}{}                                 &                      &                          &                                &                                 & Correctness & Clarity & Quality & Avg. \\ \midrule
\multirow{4}{*}{Qwen2.5-VL}  & Clean Image                                          & 0.00                 & 2.72                     & 80.34                          & 79.34                           & 4.31        & 4.88    & 4.98    & 4.72 \\
                              & VLMInferSlow                                         & 0.00                 & 5.45 (2.0$\times$)                     & 169.98 (2.1$\times$)                         & 168.98 (2.1$\times$)                          & 4.36        & 4.95    & 5.00    & 4.77 \\
                              
                              & Verbose Image                                        & 0.15                 & 20.91 (7.7$\times$)                    & 474.56 (5.9$\times$)                         & 473.56 (6.0$\times$)                          & 3.38        & 4.34    & 4.51    & 4.08 \\
                              
                              & \textbf{Hidden Tail  (Ours) }                                 & \textbf{0.72}                 & \textbf{49.39 (18.1$\times$)}                    & \textbf{1542.55 (19.2$\times$)}                        & \textbf{159.66 (2.0$\times$)}                          & 4.33        & 4.93    & 4.98    & 4.75 \\ \midrule
\multirow{4}{*}{MiMo-VL}      & Clean   Image                                        & 0.03                 & 12.87                    & 323.66                         & 322.66                          & 4.77        & 4.55    & 4.68    & 4.67 \\
                              & VLMInferSlow                                         & 0.01                 & 20.16 (1.6$\times$)                      & 482.90 (1.5$\times$)                           & 481.91 (1.5$\times$)                            & 4.78        & 4.65    & 4.74    & 4.72 \\

                              & Verbose Image                                        & 0.52                 & 62.81 (4.9$\times$)                    & 1224.15 (3.8$\times$)                        & 1219.57 (3.8$\times$)                         & 1.93        & 2.48    & 2.60    & 2.34 \\
                              
                              & \textbf{Hidden Tail  (Ours)}                                  & \textbf{0.84}                 & \textbf{73.23 (5.7$\times$)}                    & \textbf{1831.27 (5.7$\times$)}                        & \textbf{468.02 (1.5$\times$)}                          & 4.22        & 4.28    & 4.36    & 4.29 \\ \midrule
\multirow{4}{*}{Gemma3}       & Clean Image                      & 0.00                 & 14.25                    & 211.85                         & 210.85                          & 3.49        & 4.31    & 4.47    & 4.09 \\
                              & VLMInferSlow                     & 0.00                 & 20.58 (1.4$\times$)                    & 292.28 (1.4$\times$)                         & 291.27 (1.4$\times$)                          & 3.35        & 4.31    & 4.49    & 4.05 \\
                              & Verbose Image                    & 0.45                 & 79.07 (5.5$\times$)                    & 1078.86 (5.1$\times$)                        & 1043.13 (4.9$\times$)                         & 1.35        & 1.65    & 1.78    & 1.59 \\
                              & \textbf{Hidden Tail (Ours)}              & \textbf{0.68}                 & \textbf{100.04 (7.0$\times$)}                 & \textbf{1494.50 (7.1$\times$)}                        & \textbf{297.71  (1.4$\times$)}                          & 3.03        & 3.93    & 4.09    & 3.68      \\       \bottomrule                              
\end{tabular}%
\end{table*}

\subsection{Adversarial Image Generation}
\label{sec:Adversarial-Image-Generation}

Combining our three objectives poses a multi-objective optimization challenge.
A simple sum is ineffective, as the components not only operate on different scales, but also exhibit varying convergence rates.
To address this, we adopt a two-pronged strategy involving both static and dynamic weighting.
The total loss is formulated as:
\begin{equation}
\begin{aligned}
    \mathcal{L}_{\mathrm{total}} =\;& \mu_{\mathrm{sem}} \cdot \lambda_{\mathrm{sem}} \cdot \mathcal{L}_{\mathrm{sem}} \\
    &+ \mu_{\mathrm{spe}} \cdot \lambda_{\mathrm{spe}} \cdot \mathcal{L}_{\mathrm{spe}} \\
    &+ \mu_{\mathrm{eos}} \cdot \lambda_{\mathrm{eos}} \cdot \mathcal{L}_{\mathrm{eos}} ,
\end{aligned}
\label{equ:total_loss}
\end{equation}
where $\mu_{\mathrm{sem}}$, $\mu_{\mathrm{spe}}$, and $\mu_{\mathrm{eos}}$ are fixed, manually tuned scaling factors that statically balance the initial orders of magnitude, while $\lambda_{\mathrm{sem}}$, $\lambda_{\mathrm{spe}}$, and $\lambda_{\mathrm{eos}}$ are adaptive weights dynamically adjusted to balance convergence speeds.

\mypara{Dynamic Loss Weighting}
Our method for calculating the adaptive weights builds upon Dynamic Weight Averaging (DWA)~\cite{LJD19}.
The underlying intuition is to assign higher weights to loss components that are converging more slowly, thereby drawing more attention from the optimizer.

Let $\mathcal{L}_k^{(t)}$ denote the value of the $k$ loss at the current optimization step $t$, where $k \in S = \{\mathrm{sem}, \mathrm{spe}, \mathrm{eos}\}$.
We first calculate the loss ratio $r_k^{(t)}$, which measures the change rate of each loss component relative to the previous step $t-1$:
\begin{equation}
r_k^{(t)} = \frac{\mathcal{L}_k^{(t)}}{\mathcal{L}_k^{(t-1)} + \sigma},
\end{equation}
where $\sigma$ is a small constant (e.g., $10^{-8}$) to avoid division by zero.
A larger $r_k^{(t)}$ indicates slower convergence for that loss component.
The ratios are then smoothed into a probability distribution using the softmax function with a temperature parameter $\mathcal{T}$:
\begin{equation}
\tilde{\lambda}_k^{(t)} = \frac{\exp(r_k^{(t)} / \mathcal{T})}{\sum_{j\in S} \exp(r_j^{(t)} / \mathcal{T})},
\end{equation}
where a lower $\mathcal{T}$ sharpens the distribution, while a higher $\mathcal{T}$ produces more uniform weights.
To prevent any single objective from being neglected, we enforce a lower bound $\lambda_{\min}$ on each weight and re-normalize:
\begin{equation}
\lambda_k'^{(t)} = \max(\tilde{\lambda}_k^{(t)}, \lambda_{\min}),
\label{equ:min_weight}
\end{equation}
\begin{equation}
\lambda_k^{(t)} = \frac{\lambda_k'^{(t)}}{\sum_{j\in S} \lambda_j'^{(t)}}.
\end{equation}
For the initial step ($t=1$), the weights are initialized uniformly, i.e., $\lambda_k^{(1)} = \frac{1}{|S|}$ for all $k \in S$.

\mypara{Adversarial Perturbation Optimization}
Modern VLMs typically employ an integrated image processor to handle all necessary operations, such as resizing and normalization. 
To align with this common architecture, our attack directly optimizes on the feature level after the processor.
We first pass the clean image $\boldsymbol{I}_{\mathrm{clean}}$ through the image processor to obtain the clean feature tensor $\boldsymbol{x}$.
Following previous adversarial attacks~\cite{ZZMBS24, GBGXTLL24}, we then employ Projected Gradient Descent (PGD)~\cite{MMSVT18} to iteratively craft a perturbation $\boldsymbol{\delta}$, which shares the same shape as $\boldsymbol{x}$.
At each iteration, we randomly select a prompt-response pair $(\boldsymbol{p}_i,\boldsymbol{r}_i) \in \mathcal{D}$ to enhance the generalization of the adversarial image across different textual prompts.
Starting from an initial perturbation $\boldsymbol{\delta}_0$, the updated perturbation at each step $t$ is formulated as:
\begin{equation}
\begin{aligned}
    \boldsymbol{\delta}_t = \boldsymbol{\delta}_{t-1} - \alpha \cdot \mathrm{sign}(\nabla_{\boldsymbol{\delta}_{t-1}} \mathcal{L}_{\mathrm{total}}(\boldsymbol{x}_0+\boldsymbol{\delta}_{t-1})), \\ 
    s.t. ||\boldsymbol{\delta}_t||_p \le \epsilon ,\\
\end{aligned}
\end{equation}
where $\alpha$ is the step size and $\mathrm{sign}(\cdot)$ is the element-wise sign function.
$||\boldsymbol{\delta}_t||_p \le \epsilon$ ensures the perturbation constrained, where $\epsilon$ is the maximum perturbation norm allowed.
We adopt the $\ell_\infty$-norm constraint in our experiments.
After $T$ steps, the resulting perturbation $\boldsymbol{\delta}_T$ is inverted back into the pixel space and added to $\boldsymbol{I}_{\mathrm{clean}}$ to yield the final adversarial image $\boldsymbol{I}_{\mathrm{adv}}$.
Note that this inversion process may not be perfectly lossless due to numerical approximations, but we empirically find that this does not impact the attack performance (see \cref{sec:main_results}).

\section{Evaluation}

\subsection{Experimental Setup}
\label{sec:exp_setup}

\mypara{Target Models and Datasets}
We evaluate our attack on three latest open-source VLMs, including Qwen2.5-VL-7B-Instruct (Qwen2.5-VL)~\cite{Qwen2-5-VL}, MiMo-VL-7B-RL (MiMo-VL)~\cite{MiMo}, and Gemma3-4B-IT (Gemma3)~\cite{Gemma3}.
Note that MiMo-VL is a reasoning VLM that first generates a thinking process before providing an answer, which results in naturally longer and more complex baseline outputs.
Our experiments use 10 images randomly selected from the MS-COCO dataset~\cite{LMBHPRDZ14}.
For each image, we use GPT-4o~\cite{GPT4_Report} to generate 60 diverse prompts, as described in \cref{sec:Data-Preparation}.
To evaluate prompt-agnostic effectiveness, we partition the prompt set into 40 prompts for optimization and 20 for testing.
We further adopt the target VLM to generate the responses for the 40 optimization prompts to craft the adversarial image.

\mypara{Configuration}
For each image, we run the PGD algorithm for $T = 5000$ iterations with a step size of $\alpha = 1/255$.
The maximum perturbation norm is set to $\epsilon=64/255$.
The attack is configured to induce a malicious tail of length $M=1024$ by repeatedly generating the Beginning-of-Sentence (BOS) token.
For Qwen2.5-VL and MiMo-VL, the static scaling factors in total loss are configured as $\mu_{\mathrm{sem}} = 1$, $\mu_{\mathrm{sem}} = 10^3$, and $\mu_{\mathrm{eos}} = 10^4$.
For Gemma3, these parameters are set as $\mu_{\mathrm{sem}} = 1$, $\mu_{\mathrm{sem}} = 10^3$, and $\mu_{\mathrm{eos}} = 10^3$.
For the adaptive weighting mechanism, we set the minimum weight $\lambda_{\min}=0.15$ and the temperature $\mathcal{T}=2.0$.
During the evaluation, the text generation is performed using the greedy search strategy with a maximum sequence of $2048$ tokens.
Our evaluations are conducted on 2$\times$ NVIDIA A6000 48G GPUs.

\mypara{Baseline}
We evaluate our attack compared with three baselines, including the original clean image, Verbose Image~\cite{GBGXTLL24}, and VLMInferSlow~\cite{WYCWRGYHY25}.
The clean image serves as the non-adversarial baseline.
For the white-box attack, Verbose Image, we incorporate our adaptive weighting strategy for a fair comparison.
Since its loss components operate on different scales than ours, we reconfigure its static scaling factors to $\mu_{\mathrm{sem}} = 1$, $\mu_{\mathrm{spe}} = 100$, and $\mu_{\mathrm{eos}} = 10^{-4}$, while keeping all other parameters the same as our main experimental configuration.
For the black-box attack, VLMInferSlow, we adhere to the configuration detailed in the paper~\cite{WYCWRGYHY25}.
In addition, to ensure a fair evaluation of prompt-agnostic capabilities, both baseline attacks are extended using the same dataset of prompts generated for our method.

\mypara{Evaluation Metrics}
We use three metrics to measure the attack efficacy in consuming resources.
Higher values for all three metrics indicate a more effective attack.
\begin{itemize}
    \item \textbf{Attack Success Rate (ASR)} is defined as the percentage of attack attempts that successfully cause the VLM to generate up to its maximum sequence limit.
    \item \textbf{Response Latency} is to measure the average inference cost of all the testing samples.
    \item \textbf{Output length} is the total number of tokens generated, including special tokens, and serves as a direct measure of the computational payload.
\end{itemize}

To assess how well the attack hides from users, i.e., attack stealthiness, we measure two aspects of the output.
\begin{itemize}
    \item \textbf{Visible length} is the length of the response after special tokens have been removed, which is what a user actually sees.
    An ideal attack should achieve a long overall output length while maintaining a visible length comparable to that of the response generated from a clean image.
    \item \textbf{Response Quality} is to quantify the quality of this visible response.
    We employ GPT-4.1-mini~\cite{GPT4_Report} to score the user-visible part of the generated text on a scale of 1 to 5 across three dimensions: correctness, clarity, and quality.
    Higher scores indicate better stealthiness.
    The specific prompt used for this evaluation is inspired by~\cite{LZKSB25}, and detailed in the Appendix.
\end{itemize}

\begin{figure}[t]
    \centering
    \includegraphics[width=1\linewidth]{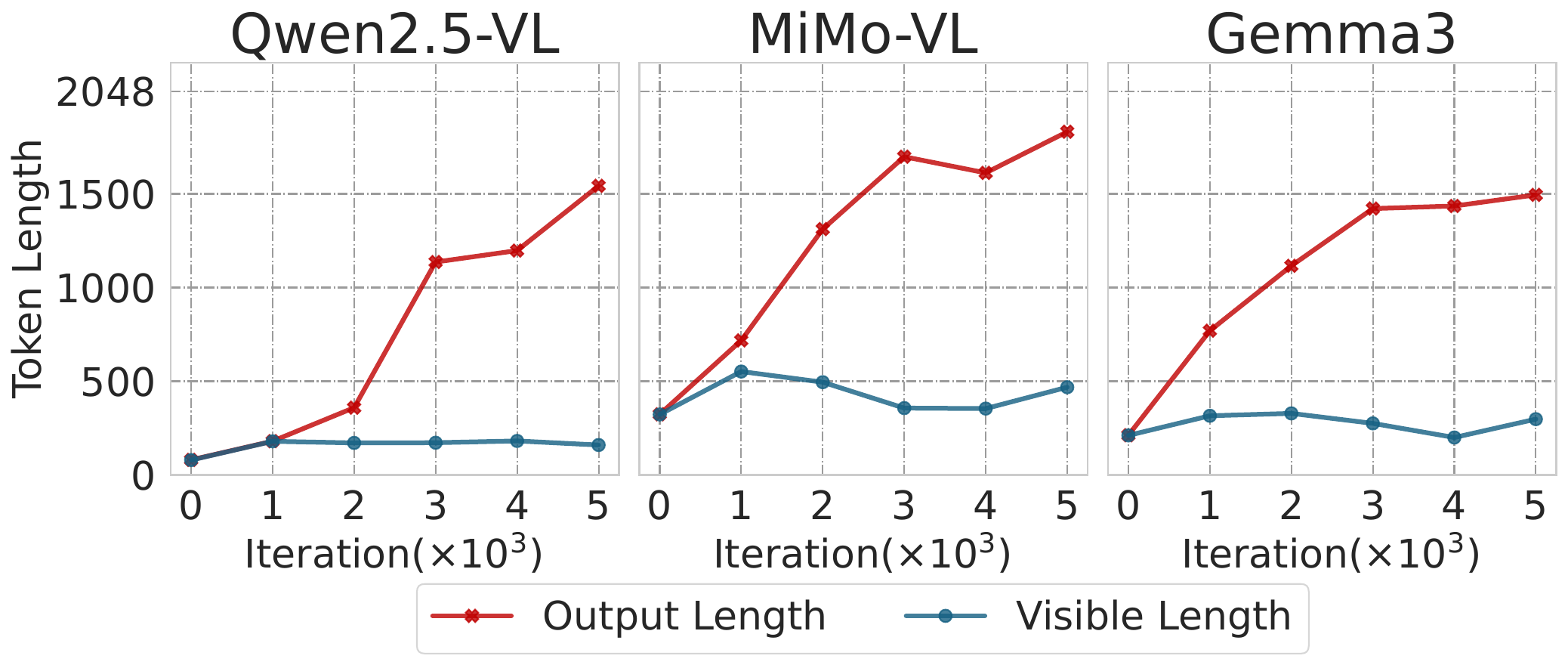}
    \caption{The effect of optimization iterations $T$ on attack performance.}
    \label{fig:iteration}
\end{figure}

\subsection{Main Results}
\label{sec:main_results}

We provide the attack performance of Verbose Image, VLMInferSlow, and our \textit{Hidden Tail} attack in \cref{tab:main}, using the performance on clean images as a non-adversarial baseline.
The VLMInferSlow attack exhibits the weakest performance.
It only marginally increases the output length on three VLMs, likely because its black-box nature prevents accurate gradient estimation for effective optimization.
The Verbose Image attack shows partial effectiveness, achieving longer average output lengths, e.g., 474.56 on Qwen2.5-VL, 1224.15 on MiMo-VL, and 1078.86 on Gemma3.
However, the output tokens are visible to users, which contain much unrelated and repeated content.
Thus, the average response quality shows a significant decrease compared with clean images, compromising attack stealthiness.
This outcome is expected, as Verbose Image is designed solely to increase inference cost without considering stealthiness, but its overall effectiveness remains limited compared to our attack.
Our attack achieves the best of both aspects.
It demonstrates remarkable attack effectiveness, with an ASR of 0.72 on Qwen2.5-VL, 0.84 on MiMo-VL, ang 0.68 on Gemma3.
This translates to average output lengths of 1542.6, 1831.3, and 1494.50, respectively, achieving an increase of up to 19.2$\times$, 5.7$\times$, and 7.1$\times$ over the baseline of clean images.
Crucially, this is achieved while keeping the visible length comparable to that of clean images, thus preserving response quality and stealthiness.
The results demonstrate that our attack significantly outperforms existing methods, achieving superior attack effectiveness while maintaining stealthiness.
We also provide the output distributions analysis in the Appendix for more details.

\subsection{Ablation Study}

To understand the influence of the key components of our attack, we conduct a comprehensive ablation study.
We first analyze several factors in our optimization hyperparameters, including the optimization iteration $T$ and the perturbation norm $\epsilon$.
Then we investigate the attack objective design, analyzing the contribution of each individual loss term and the impact of different choices for the target special token.
We further analyze the influence of inference settings, including the sampling strategy and maximum token limit, provided in the Appendix.
Unless otherwise specified, all experiments adhere to the default setup described in \cref{sec:exp_setup}, and for clarity, we only report the average Response Quality score as a summary metric.

\mypara{Effect of Iteration $T$}
We study how the number of optimization iterations $T$ affects the attack performance.
As illustrated in \cref{fig:iteration}, the average output length improves as $T$ increases.
Conversely, the visible length remains consistently low and stable across all iteration counts.
This shows that using more iterations leads to a stronger attack, but it also requires more time and computational resources to generate the adversarial image.

\begin{figure}[t]
    \centering
    \includegraphics[width=1\linewidth]{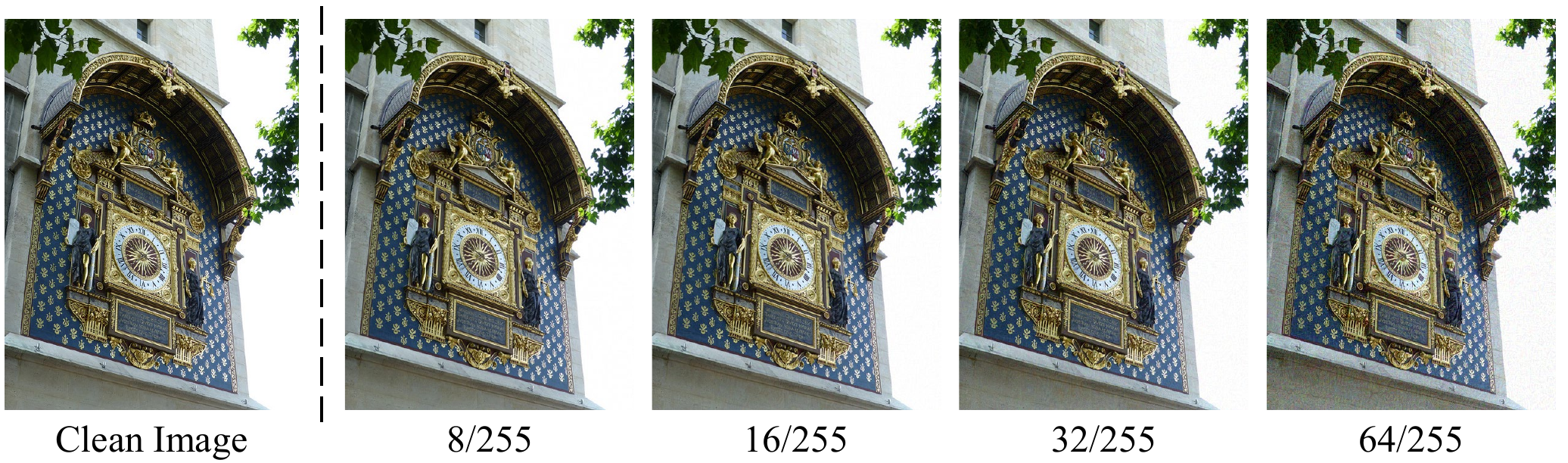}
    \caption{Examples of \textit{Hidden Tail} attack using different perturbation norms $\epsilon$.}
    \label{fig:epsilon}
\end{figure}

\mypara{Effect of Perturbation Norm $\epsilon$}
The perturbation norm $\epsilon$ determines the maximum allowed change applied to the image features.
We evaluate how different values of $\epsilon$, ranging from $8/255$ to $64/255$, affect the attack performance.
As shown in \cref{tab:epsilon}, a larger $\epsilon$ generally leads to a more effective attack.
More importantly, the example images in \cref{fig:epsilon} demonstrate that even with $\epsilon=64/255$, the changes to the final image are not conspicuous.
This high degree of stealth is a key benefit of our feature-level optimization.
When the perturbation is converted from the feature space back to pixels, the changes are distributed subtly across the entire image, making them difficult for users to notice.

\begin{table}[t]
\centering
\caption{Evaluation results of \textit{Hidden Tail} Attack on Qwen2.5-VL under varying perturbation norm $\epsilon$.}
\label{tab:epsilon}
\small
\setlength{\tabcolsep}{6pt}
\begin{tabular}{c | ccc  c cccc}
\toprule
$\epsilon$ & ASR & Latency & \makecell{Output \\ Length} & \makecell{Visible \\ Length} & \makecell{Response \\ Quality} \\ \midrule
8/255      & 0.00             & 5.33                 & 163.99                                   & 162.98                                      & 4.65                          \\
16/255     & 0.29             & 24.07                & 748.44                                   & 307.11                                      & 4.59                          \\
32/255     & 0.44             & 32.84                & 1008.45                                  & 160.45                                      & 4.60                          \\
64/255     & 0.72             & 49.39                & 1542.55                                  & 159.66                                      & 4.75 \\ \bottomrule
\end{tabular}%
\end{table}

\begin{table}[t]
\centering
\caption{Evaluation results of \textit{Hidden Tail} Attack on Qwen2.5-VL under varying special tokens.}
\label{tab:special_tokens}
\small
\setlength{\tabcolsep}{2pt}
\begin{tabular}{l|cccccccc}
\toprule
Special  Token & ASR  & Latency  & \makecell{Output \\ Length} & \makecell{Visible \\ Length}  & \makecell{Response \\ Quality} \\  \midrule
Clean Image                               & 0.00                               & 2.72                                   & 80.34                          & 79.34                           & 4.72    \\ \midrule
$\mathtt{im\_start}$                              & 0.72                               & 49.39                                  & 1542.55                        & 159.66                             & 4.75    \\
$\mathtt{object\_ref\_start}$                      & 0.84                               & 61.82                                  & 1755.61                        & 197.77                       & 4.63    \\
$\mathtt{object\_ref\_end}$                        & 0.70                               & 57.20                                  & 1637.14                        & 138.63                          & 4.56    \\
$\mathtt{box\_start}$                             & 0.85                               & 60.08                                  & 1762.24                        & 158.43                          & 4.69    \\
$\mathtt{box\_end}$                               & 0.84                               & 61.65                                  & 1778.71                        & 197.14                          & 4.75    \\
$\mathtt{quad\_start}$                            & 0.89                               & 63.47                                  & 1844.07                        & 152.38                         & 4.75    \\
$\mathtt{quad\_end}$                              & 0.81                               & 56.99                                  & 1682.52                        & 150.15                          & 4.72   \\ \midrule
Average                                      & 0.81                               & 58.66                                  & 1714.69                        & 164.88                          & 4.69    \\ \bottomrule
\end{tabular}%
\end{table}

\begin{figure}[t]
    \centering
    \includegraphics[width=0.7\linewidth]{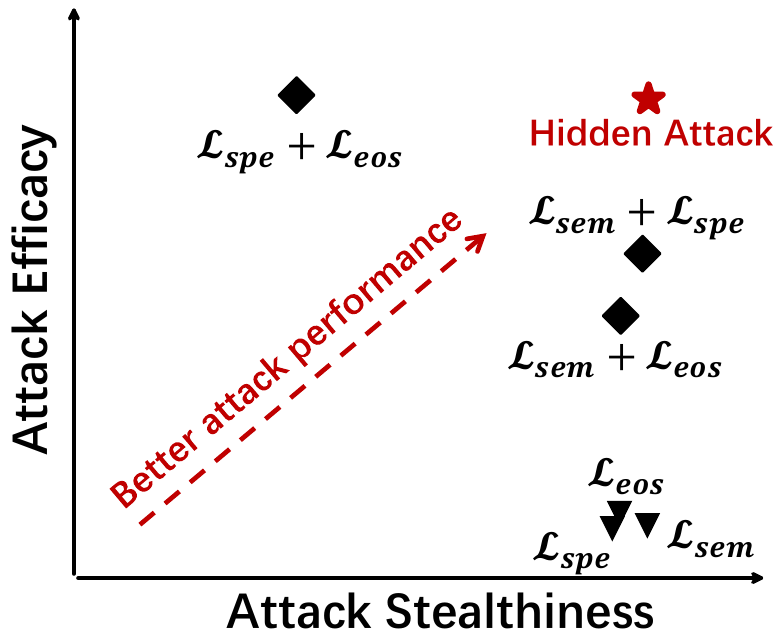}
    \caption{Comparison of different combinations of the three loss objectives. The target VLM is Qwen2.5-VL. }
    \label{fig:loss}
\end{figure}

\mypara{Effect of Loss Objectives}
To validate the contribution of each component in our loss function, we conduct an ablation study on all possible combinations.
We adopt Qwen2.5-VL as the target VLM, and use output length and response quality as the primary metrics for attack efficacy and stealthiness, respectively.
As shown in \cref{fig:loss}, the results demonstrate that no single objective is sufficient to conduct a successful attack.
Moreover, any combination of just two objectives exhibits critical limitations, either by failing to generate a substantial output length or by severely compromising the attack stealthiness. 
These findings confirm that all three of our loss components are necessary, with each playing a distinct and synergistic role in achieving the effective and stealthy attack.
We also present the detailed evaluation results and analysis in the Appendix.

\mypara{Special Token Selection}
To verify that our attack is not solely dependent on the BOS token, we adopt six additional special tokens native to the Qwen2.5-VL tokenizer for the hidden tail induction objective.
As presented in~\cref{tab:special_tokens}, all tested special tokens achieve performance comparable to the BOS token.
On average, the output length reaches 1714.69, while the visible length and response quality are maintained at 164.9 and 4.69, respectively.
This consistency suggests that the vulnerability is not specific to a single special token but rather reflects a more fundamental issue.
Our findings indicate that special tokens in VLMs are broadly susceptible to hidden tail attacks, raising serious concerns about token-level security.
The details of these tokens and more results are provided in the Appendix.

\section{Discussion}

\begin{figure}
    \centering
    \includegraphics[width=0.95\linewidth]{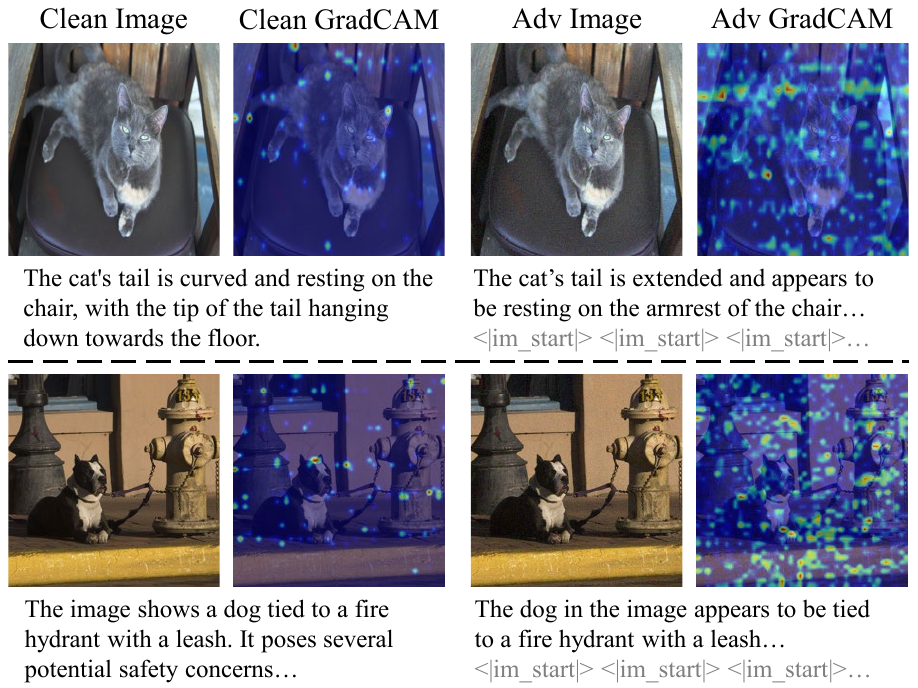}
    \caption{GradCAM visualization of clean and adversarial images. The target VLM is Qwen2.5-VL. The question for the first example is \textit{how is the cat's tail positioned?} and the second is \textit{are there any safety concerns for the dog?}}
    \label{fig:GradCAM}
\end{figure}

\mypara{Visual Interpretation}
To better understand our attack from the model perspective, we employ GradCAM~\cite{SCDVPB17} to visualize the VLM's attention map on the input image during generation.
As presented in \cref{fig:GradCAM}, the resulting attention maps reveal a significant contrast between the clean and adversarial images.
For the clean image, the attention map is sparse, which we attribute to the model relying more heavily on the text prompt.
In contrast, the attention map for the adversarial image is both significantly more intense and broadly diffused across the entire visual field.
We speculate that the adversarial perturbation induces a state of attentional confusion, causing the visual focus to scatter across the image and thereby enabling the non-terminating, long-tail output.

\mypara{Transferability}
We investigated the transferability of our attack by applying adversarial images crafted for Qwen2.5-VL to other VLMs.
This includes VLMs from the same family but different scales, i.e., the 3B and 32B versions of Qwen2.5-VL~\cite{Qwen2-5-VL}, and VLMs from entirely different developers, i.e., MiMo-VL~\cite{MiMo}, Gemma3-4B-IT~\cite{Gemma3}, and Phi-Vision-Instruct~\cite{Phi3}.
The results, which are detailed in the Appendix, consistently demonstrate that our attack exhibits little transferability.
We hypothesize that this stems from our attack's reliance on token-level optimization.
The internal representations and handling of special tokens are highly specific to each individual VLM, making the adversarial features non-transferable.
This inherent non-transferability makes the attack highly targeted, allowing an attacker to affect specific VLM services without causing unintended collateral impact on unrelated models.

\section{Conclusion}

In this work, we propose \textit{Hidden Tail}, a stealthy and effective prompt-agnostic resource consumption attack against VLMs. 
Our method crafts adversarial images using a multi-objective optimization that forces a VLM into a non-terminating loop of generating user-invisible special tokens, while ensuring the initial visible response remains semantically consistent.
Extensive experiments demonstrate that \textit{Hidden Tail} increases output length by up to 19.2$\times$, reaching the maximum token limit while preserving the attack stealthiness. 
This highlights a fundamental vulnerability in how VLMs handle generative control tokens, a threat that is amplified as model output capacities grow.

%-------------------------------------------------------------------------------
\begin{small}
\bibliographystyle{plain}
\bibliography{reference}
\end{small}
%-------------------------------------------------------------------------------

\newpage
\appendix
\onecolumn

\section{Related Work}

\subsection{Vision-Language Models}

Vision-Language Models (VLMs) are multimodal models that can jointly process visual and textual modalities to perform multimodal tasks such as image captioning~\cite{SCBCHF22} and visual question answering (VQA)~\cite{LLYYLZ23}.
Early VLMs, such as BLIP~\cite{LLXH22}, InstructBLIP~\cite{DLLTZWLFH23}, and MiniGPT-4~\cite{ZCSLE23}, introduce image encoders (e.g., Vision Transformers) paired with large language models (LLMs) as decoders.
These models use query-based feature projection to translate visual features into prompts that LLMs can process, achieving strong performance in open-ended tasks.
With the rapid development of powerful LLMs, the capability of VLMs has significantly expanded.
Models like Qwen2.5-VL~\cite{Qwen2-5-VL}, Gemma-3~\cite{Gemma3}, and Phi-3.5-vision~\cite{Phi3}, adopt more tightly coupled vision-language architectures, enabling unified reasoning across modalities and supporting complex tasks such as chart understanding~\cite{HCFQZJCJ24} and visual programming~\cite{GK23}.
Despite their impressive capabilities, VLMs inherit the sensitivity to input perturbations from both vision and language components.
This creates new, underexplored attack surfaces, particularly in how image features influence the generation process.
Our work focuses on a novel class of attacks that exploit the adversarial image against VLMs to induce resource consumption.

\section{Details of Data Preparation}
To create a diverse dataset of prompt-response pairs for each clean image, we employ an iterative, multi-turn querying strategy to query GPT-4o.
Given a clean image, we first conduct an initial query to generate a baseline set of prompts.
In each subsequent turn, we refine the query to explicitly request new prompts that are semantically distinct from all previously generated ones.
This process is repeated for a predefined number of rounds to accumulate a rich and varied set of prompts.
Our adopted prompts are shown below.

\begin{tcolorbox}[
breakable,
title=Prompt for Data Preparation, 
boxrule=0pt,
sharp corners,
enhanced,
]
\# Initial Query

Generate 10 different questions based on this image. 
\\

\# Subsequent Query 

Generate 10 more questions for this image, ensuring they are different from the previous set.
\end{tcolorbox}

\section{Details of Response Quality Evaluation}
We use the following prompt to query GPT-4.1-mini~\cite{GPT4_Report} to score the user-visible part for response quality evaluation.

\begin{tcolorbox}[
breakable,
title=Prompt for Text Quality Evaluation, 
boxrule=0pt,
sharp corners,
enhanced,
]
Given: \\

- A question: \{question\}

- An answer: \{answer\}
\\

\#\#\# Your task: 
1. Evaluate the answer based on the following criteria:

- **Answer Correctness:** Does the answer accurately and fully answer the question based on the provided image? 

- **Clarity \& Readability:** Is the writing of the answer clear, coherent, easy to understand?

- **Text Quality:** Is the writing free of major grammatical errors?
\\

2. Scoring standards for each criterion (Important: All scores must be integers from 1 to 5.):

- **5:** Excellent. Fully meets expectations. No major weaknesses. 

- **4:** Good. Minor weaknesses that do not seriously impact quality. 

- **3:** Fair. Some noticeable issues that reduce effectiveness.

- **2:** Poor. Serious flaws or missing key aspects.

- **1:** Very poor. Does not meet basic expectations.
\\

3. For the criterion, provide: 

- A score (from 1 to 5) 

- An explanation of why you gave this score.
\\

4. Output your evaluation in the following JSON format: 

\{ ``answer\_correctness\_score": X, 

``answer\_correctness\_explanation": ``...",

``clarity\_readability\_score": X, 

``clarity\_readability\_explanation": ``...",

``text\_quality\_score": X 

``text\_quality\_explanation": ``..." \}

\end{tcolorbox}

\section{Additional Results}

\begin{figure}
    \centering
    \includegraphics[width=0.8\linewidth]{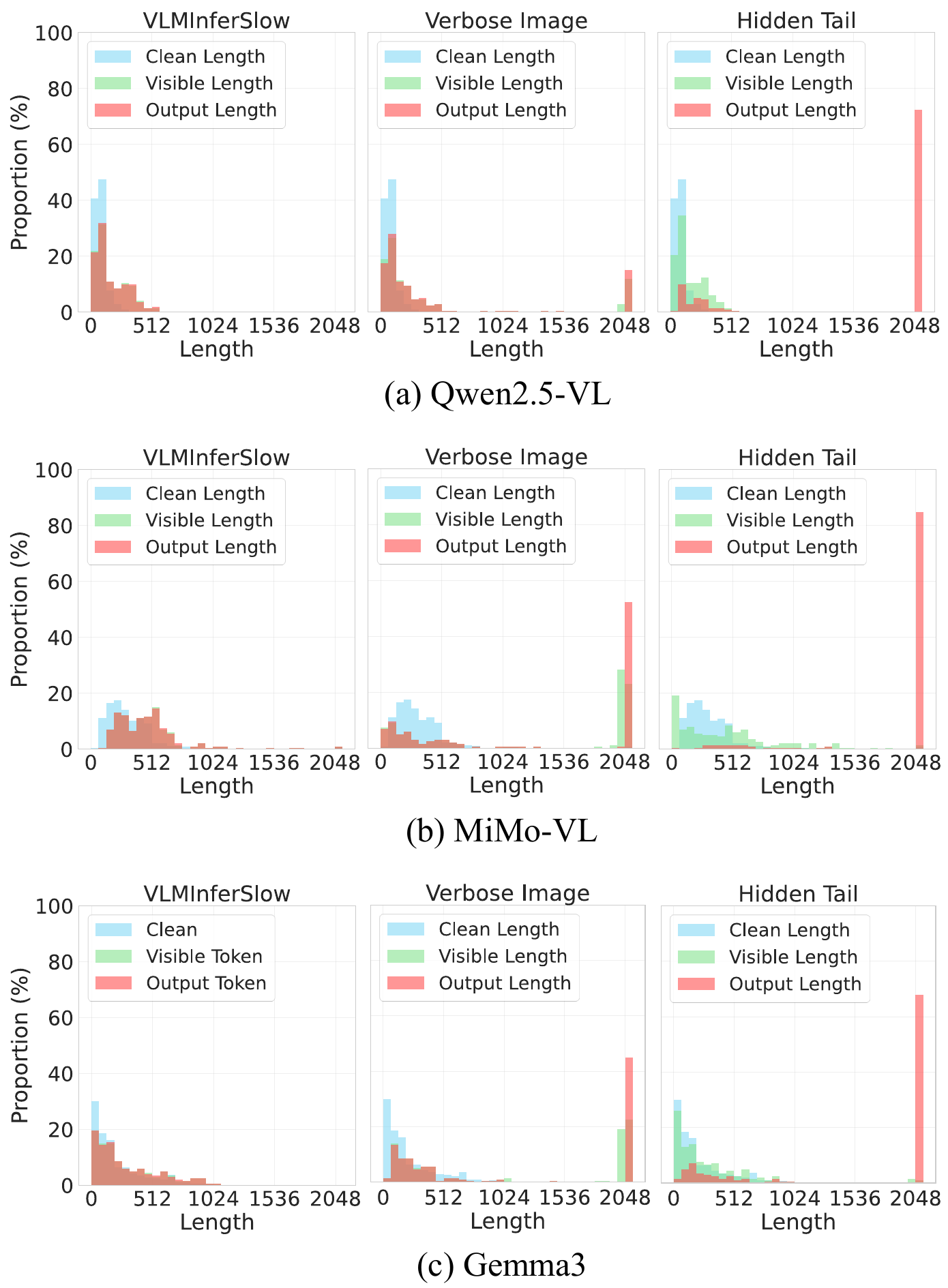}
    \caption{A comparison of output length distributions for each attack method on all the target VLMs. Each subplot visualizes three distinct distributions: (1) the output length from the clean image (baseline), (2) the visible length of the adversarial output, and (3) the total output length of the adversarial output.}
    \label{fig:Distribution}
\end{figure}

\begin{table*}[t]
\centering
\caption{Evaluation results of \textit{Hidden Tail} Attack on Qwen2.5-VL and MiMo-VL under varying perturbation norm $\epsilon$.}
\label{tab:epsilon_detail}
\resizebox{1\textwidth}{!}{
\begin{tabular}{c c | ccc  c cccc}
\toprule
\multirow{2}{*}{Model}      & \multirow{2}{*}{$\epsilon$} & \multirow{2}{*}{ASR ($\uparrow$)} & \multirow{2}{*}{Latency ($\uparrow$)} & \multirow{2}{*}{\makecell{Output \\ Length} ($\uparrow$)} & \multirow{2}{*}{\makecell{Visible \\ Length} ($\downarrow$)} & \multicolumn{4}{c}{Response Quality ($\uparrow$)}      \\ \cmidrule{7-10} 
                            &                           &                                &                          &                      &                                 & Correctness & Clarity & Quality & Avg. \\ \midrule
\multirow{4}{*}{Qwen2.5-VL}     & 8/255                     & 0.00                 & 5.33                     & 163.99                         & 162.98                          & 4.09        & 4.88    & 4.98    & 4.65 \\
                                & 16/255                    & 0.29                 & 24.07                    & 748.44                         & 307.11                          & 4.16        & 4.78    & 4.83    & 4.59 \\
                                & 32/255                    & 0.44                 & 32.84                    & 1008.45                        & 160.45                          & 4.01        & 4.84    & 4.96    & 4.60 \\
                                & 64/255                    & 0.72                 & 49.39                    & 1542.55                        & 159.66                          & 4.33        & 4.93    & 4.98    & 4.75 \\ \midrule
\multirow{4}{*}{MiMo-VL}        & 8/255                     & 0.01                 & 23.92                    & 604.37                         & 603.38                          & 4.19        & 4.47    & 4.63    & 4.43 \\
                                & 16/255                    & 0.66                 & 57.01                    & 1486.34                        & 435.54                          & 4.50        & 4.54    & 4.63    & 4.56 \\
                                & 32/255                    & 0.73                 & 63.61                    & 1654.97                        & 477.22                          & 4.54        & 4.68    & 4.75    & 4.66 \\
                                & 64/255                    & 0.84                 & 73.23                    & 1831.27                        & 468.02                          & 4.22        & 4.28    & 4.36    & 4.29 \\ \midrule
\multirow{4}{*}{Gemma3}         & 8/255                     & 0.00                 & 20.979                   & 289.82                         & 288.82                          & 3.15        & 4.26    & 4.46    & 3.96 \\
                                & 16/255                    & 0.03                 & 24.966                   & 343.11                         & 332.95                          & 2.84        & 4.14    & 4.42    & 3.80 \\
                                & 32/255                    & 0.55                 & 85.711                   & 1254.67                        & 394.01                          & 2.18        & 3.34    & 3.57    & 3.03 \\
                                & 64/255                    & 0.68                 & 100.039                  & 1494.5                         & 297.71                          & 3.03        & 3.93    & 4.09    & 3.68 \\
                                \bottomrule
\end{tabular}%
}
\end{table*}

\begin{table*}[t]
\centering
\caption{Attack performance of different combinations of the three loss objectives. The target VLM is Qwen2.5-VL.}
\label{tab:loss}
\resizebox{1\textwidth}{!}{
\begin{tabular}{ccc|cccccccc}
\toprule
\multirow{2}{*}{\textbf{$\mathcal{L}_{sem}$}} & \multirow{2}{*}{\textbf{$\mathcal{L}_{spe}$}} & \multirow{2}{*}{\textbf{$\mathcal{L}_{eos}$}} & \multirow{2}{*}{ASR  ($\uparrow$)} & \multirow{2}{*}{Latency  ($\uparrow$)} & \multirow{2}{*}{\makecell{Output \\ Length} ($\uparrow$)} & \multirow{2}{*}{\makecell{Visible \\ Length} ($\downarrow$)} & \multicolumn{4}{c}{Response Quality    ($\uparrow$)} \\ \cmidrule{8-11} 
                                              &                                               &                                               &                                    &                                        &                                &                                 & Correctness     & Clarity     & Quality    & Avg.    \\ \midrule
\Checkmark                                    &                                               &                                               & 0.00                               & 5.52                                   & 167.93                         & 166.93                          & 4.29            & 4.94        & 4.99       & 4.74    \\
\multicolumn{1}{l}{}                          & \Checkmark                                    & \multicolumn{1}{l|}{}                         & 0.00                               & 5.08                                   & 158.58                         & 157.58                          & 3.64            & 4.75        & 4.97       & 4.45    \\
\multicolumn{1}{l}{}                          & \multicolumn{1}{l}{}                          & \Checkmark                                    & 0.00                               & 6.85                                   & 207.41                         & 206.41                          & 3.82            & 4.76        & 4.94       & 4.51    \\
\Checkmark                                    & \Checkmark                                    & \multicolumn{1}{l|}{}                         & 0.47                               & 36.78                                  & 1044.83                        & 135.09                          & 4.27            & 4.87        & 4.95       & 4.70    \\
\Checkmark                                    & \multicolumn{1}{l}{}                          & \Checkmark                                    & 0.34                               & 27.48                                  & 844.51                         & 830.00                          & 4.34            & 4.63        & 4.59       & 4.52    \\
\multicolumn{1}{l}{}                          & \Checkmark                                    & \Checkmark                                    & 0.72                               & 51.10                                  & 1555.38                        & 52.94                           & 1.57            & 1.94        & 2.01       & 1.84    \\ \midrule
\Checkmark                                    & \Checkmark                                    & \Checkmark                                    & 0.72                               & 49.39                                  & 1542.55                        & 159.66                          & 4.33            & 4.93        & 4.98       & 4.75    \\ \bottomrule
\end{tabular}%
}
\end{table*}

\begin{table*}[t]
\centering
\caption{Evaluation results on different special tokens. The target VLM is Qwen2.5-VL.}
\label{tab:special_tokens_detail}
\resizebox{1\textwidth}{!}{
\begin{tabular}{l|cccccccc}
\toprule
\multirow{2}{*}{Special   Token} & \multirow{2}{*}{ASR  ($\uparrow$)} & \multirow{2}{*}{Latency  ($\uparrow$)} & \multirow{2}{*}{\makecell{Output \\ Length} ($\uparrow$)} & \multirow{2}{*}{\makecell{Visible \\ Length} ($\downarrow$)} & \multicolumn{4}{c}{Response Quality    ($\uparrow$)} \\\cmidrule{6-9}
                                          &                                    &                                        &                                &                                 & Correctness     & Clarity     & Quality    & Avg.    \\ \midrule
Clean Image                               & 0.00                               & 2.72                                   & 80.34                          & 79.34                           & 4.31            & 4.88        & 4.98       & 4.72    \\ \midrule
$<|\mathtt{im\_start}|>$                              & 0.72                               & 49.39                                  & 1542.55                        & 159.66                          & 4.33            & 4.93        & 4.98       & 4.75    \\
$<|\mathtt{object\_ref\_start}|>$                      & 0.84                               & 61.82                                  & 1755.61                        & 197.77                          & 4.17            & 4.83        & 4.89       & 4.63    \\
$<|\mathtt{object\_ref\_end}|>$                        & 0.70                               & 57.20                                  & 1637.14                        & 138.63                          & 4.13            & 4.72        & 4.82       & 4.56    \\
$<|\mathtt{box\_start}|>$                             & 0.85                               & 60.08                                  & 1762.24                        & 158.43                          & 4.21            & 4.91        & 4.95       & 4.69    \\
$<|\mathtt{box\_end}|>$                               & 0.84                               & 61.65                                  & 1778.71                        & 197.14                          & 4.34            & 4.92        & 5.00       & 4.75    \\
$<|\mathtt{quad\_start}|>$                            & 0.89                               & 63.47                                  & 1844.07                        & 152.38                          & 4.33            & 4.93        & 5.00       & 4.75    \\
$<|\mathtt{quad\_end}|>$                              & 0.81                               & 56.99                                  & 1682.52                        & 150.15                          & 4.34            & 4.88        & 4.95       & 4.72   \\ \midrule
Average                                      & 0.81                               & 58.66                                  & 1714.69                        & 164.88                          & 4.26            & 4.87        & 4.94       & 4.69    \\ \bottomrule
\end{tabular}%
}
\end{table*}

\subsection{Length Distribution Analysis}
A more detailed analysis of the output length distributions in \cref{fig:Distribution} further validates our attack superiority.
For a clean image, the actual output length is nearly identical to the visible length, differing only by a single EOS token.
This clean length serves as the baseline for a normal, user-perceived response.
An ideal stealth attack is to break the correlation of visible and actual output length.
The visible length distribution should closely match that of clean images, while the actual output length distribution should be heavily concentrated near the maximum limit (2048).
The baseline methods, however, fail to meet this objective.
The output distributions for VLMInferSlow barely deviate from the clean baseline, confirming its ineffectiveness.
For Verbose Image, the distributions of output length and visible length are nearly identical.
While it successfully generates a long output, the visible response is also long, making the attack immediately obvious to a user.
Our \textit{Hidden Tail} attack perfectly matches the ideal pattern that the visible length and the clean length share a similar distribution.
Meanwhile, its actual output length distribution shows that most samples reach the maximum token limit, demonstrating that our method can secretly and effectively exhaust the resources of the target VLM service.

\subsection{Effect of Perturbation Norm $\epsilon$}

We provide supplementary evaluation results in \cref{tab:epsilon_detail}.
The results show that MiMo-VL and Gemma3 exhibit a similar trend to Qwen2.5-VL.
A larger perturbation norm $\epsilon$ generally leads to a more effective attack.

\subsection{Loss Objective Design}

To validate the contribution of each component in our composite loss function, we conduct an ablation study on all possible combinations of the loss terms.
We evaluate each combination on the Qwen2.5-VL model.
The results are summarized in \cref{tab:loss}.
Our findings confirm that no single objective is sufficient to mount a successful attack. 
More importantly, combinations of two objectives each exhibit critical weaknesses. 
While the combination of $\mathcal{L}_{\mathrm{sem}}+\mathcal{L}_{\mathrm{eos}}$ can produce long outputs, the tail consists of semantically incoherent text rather than a controlled token sequence. 
Conversely, $\mathcal{L}_{\mathrm{spe}}+\mathcal{L}_{\mathrm{eos}}$ combining fails to preserve the initial response quality, thus sacrificing stealthiness.
Finally, the combination $\mathcal{L}_{\mathrm{sem}}+\mathcal{L}_{\mathrm{eos}}$ is ineffective at generating a long tail, as it lacks the crucial EOS suppression mechanism.
These results lead to a clear conclusion: all three of our proposed loss objectives are necessary and play distinct, complementary roles.
$\mathcal{L}_{\mathrm{sem}}$ ensures stealth, $\mathcal{L}_{\mathrm{spe}}$ creates the specific attack payload, and $\mathcal{L}_{\mathrm{eos}}$ sustains the generation process.
Only in concert can they produce the desired, effective, and stealthy long-response attack.

\begin{table*}[t]
\centering
\caption{Evaluation results when using Nucleus Sampling as the sampling strategy.}
\label{tab:strategy}
\resizebox{1\textwidth}{!}{
\begin{tabular}{cccccccccc}
\toprule
\multirow{2}{*}{Model} & \multirow{2}{*}{Strategy} & \multirow{2}{*}{ASR ($\uparrow$)} & \multirow{2}{*}{Latency ($\uparrow$)} & \multirow{2}{*}{\makecell{Output \\ Length} ($\uparrow$)} & \multirow{2}{*}{\makecell{Visible \\ Length} ($\downarrow$)} & \multicolumn{4}{c}{Response Quality ($\uparrow$)} \\ \cmidrule{7-10} 
                                &                           &                                    &                                        &                                &                                 & Correctness     & Clarity     & Quality    & Avg.    \\ \midrule
\multirow{2}{*}{Qwen2.5-VL}       & Greedy Search             & 0.72                               & 49.39                                  & 1542.55                        & 159.66                          & 4.33            & 4.93        & 4.98       & 4.75    \\
                                & Nucleus Sampling                 & 0.45                               & 38.87                                  & 1076.26                        & 142.11                          & 4.10            & 4.69        & 4.73       & 4.51    \\ \midrule
\multirow{2}{*}{MiMo-VL}           & Greedy Search             & 0.84                               & 73.23                                  & 1831.27                        & 468.02                          & 4.22            & 4.28        & 4.36       & 4.29    \\
                                & Nucleus Sampling                 & 0.68                               & 57.78                                  & 1525.99                        & 305.30                          & 4.32            & 4.34        & 4.42       & 4.36   \\ \midrule
\multirow{2}{*}{Gemma3}         & Greedy Search             & 0.68                               & 100.04                                 & 1494.50                        & 297.71                          & 3.03            & 3.93        & 4.09       & 3.68    \\
                                & Nucleus Sampling                 & 0.64                               & 102.14                                 & 1427.14                        & 269.17                          & 2.95            & 3.93        & 4.04       & 3.64 \\
                                \bottomrule
\end{tabular}%
}
\end{table*}

\begin{table*}[t]
\centering
\caption{Evaluation results when configuring max output length to 4096.}
\label{tab:4096}
\resizebox{1\textwidth}{!}{
\begin{tabular}{cccccccccc}
\toprule
\multirow{2}{*}{Model} & \multirow{2}{*}{Max Length} & \multirow{2}{*}{ASR  ($\uparrow$)} & \multirow{2}{*}{Latency  ($\uparrow$)} & \multirow{2}{*}{\makecell{Output \\ Length} ($\uparrow$)} & \multirow{2}{*}{\makecell{Visible \\ Length} ($\downarrow$)} & \multicolumn{4}{c}{Response Quality    ($\uparrow$)} \\ \cmidrule{7-10} 
                                &                            &                                    &                                        &                                &                                 & Correctness     & Clarity     & Quality    & Avg.    \\ \midrule
\multirow{2}{*}{Qwen2.5-VL}       & 2048                       & 0.72                               & 49.39                                  & 1542.55                        & 159.66                          & 4.33            & 4.93        & 4.98       & 4.75    \\
                                & 4096                       & 0.52                               & 117.098                                & 2365.27                        & 143.8                           & 4.15            & 4.84        & 4.95       & 4.65    \\ \midrule
\multirow{2}{*}{MiMo-VL}           & 2048                       & 0.84                               & 73.23                                  & 1831.27                        & 468.02                          & 4.22            & 4.28        & 4.36       & 4.29    \\
                                & 4096                       & 0.7                                & 150.53                                 & 3010.68                        & 349.36                          & 4.52            & 4.48        & 4.53       & 4.51    \\ \midrule
\multirow{2}{*}{Gemma3}          & 2048                       & 0.68                               & 100.04                                 & 1494.50                        & 297.71                          & 3.03            & 3.93        & 4.09       & 3.68    \\
                                & 4096                       & 0.63                               & 191.44                                 & 2713.93                        & 324.06                          & 3.06            & 3.71        & 3.81       & 3.53     \\
                                \bottomrule
\end{tabular}%
}
\end{table*}

\begin{table*}[!t]
\centering
\caption{Evaluation results of attack transferability. The target VLM is Qwen2.5-VL (7B version).}
\label{tab:transferability}
\resizebox{1\textwidth}{!}{
\begin{tabular}{l|cccccccc}
\toprule
\multirow{2}{*}{Target VLM} & \multirow{2}{*}{ASR  ($\uparrow$)} & \multirow{2}{*}{Latency  ($\uparrow$)} & \multirow{2}{*}{\makecell{Output \\ Length} ($\uparrow$)} & \multirow{2}{*}{\makecell{Visible \\ Length} ($\downarrow$)} & \multicolumn{4}{c}{Response Quality    ($\uparrow$)} \\
                                          &                                    &                                        &                                &                                 & Correctness     & Clarity     & Quality    & Avg.    \\ \midrule
Qwen2.5-VL-3B-Instruct                                     & 0.00                               & 3.72                                   & 94.67                          & 93.68                           & 3.89            & 4.86        & 4.96       & 4.57    \\
Qwen2.5-VL-32B-Instruct                                    & 0.00                               & 636.95                                 & 212.52                         & 211.52                          & 4.20            & 4.91        & 4.99       & 4.70    \\ \midrule
MiMo-VL-7B-RL                                              & 0.00                               & 13.06                                  & 330.55                         & 329.55                          & 4.72            & 4.64        & 4.71       & 4.69    \\
Gemma3-4B-IT                                              & 0.00                               & 18.54                                  & 291.19                         & 290.18                          & 3.93            & 4.87        & 4.99       & 4.60    \\
Phi3.5-Vision-Instruct                                    & 0.00                               & 3.36                                   & 41.48                          & 39.48                           & 4.21            & 4.94        & 5.00       & 4.72 \\ \bottomrule
\end{tabular}%
}
\end{table*}

\subsection{Special Token Selection}
To verify that our attack is not contingent on a specific choice of special token, we evaluate its performance using several different tokens for the hidden tail induction objective.
On the Qwen2.5-VL, we test a set of six distinct special tokens native to its tokenizer.
The six special tokens selected from the Qwen2.5-VL tokenizer, along with their intended functions, are detailed below:
\begin{itemize}
    \item  $<|\mathtt{object\_ref\_start}|>$ and $<|\mathtt{object\_ref\_end}|>$: These tokens demarcate an object reference, a mechanism used to align a specific image region with its corresponding textual description.
    \item  $<|\mathtt{box\_start}|>$ and $<|\mathtt{box\_end}|>$: These tokens delimit the coordinates of a rectangular bounding box, which specifies the precise 2D location and scale of an object within the image.
    \item  $<|\mathtt{quad\_start}|>$ and $<|\mathtt{quad\_end}|>$: These tokens delimit the coordinates for a quadrilateral area, allowing for more flexible localization of non-rectangular objects, such as skewed text blocks for Optical Character Recognition (OCR).
\end{itemize}
The results summarized in \cref{tab:special_tokens_detail} demonstrate that the attack remains highly effective regardless of the specific token targeted.
On average, across all tested tokens, the attack achieves a substantial output length of 1714.7, while the visible length and response quality are maintained at 164.9 and 4.69, respectively.
This robustness indicates that the vulnerability is not an artifact of a single token but is more fundamental.
Our findings suggest that numerous special tokens within VLMs are susceptible to this hidden tail manipulation, raising broader concerns about token-level security in these architectures.

\subsection{Effect of Sampling Strategies}
We also evaluate how different sampling strategies affect our attack performance.
In addition to our default method, Greedy Search, we evaluate the attack using Nucleus Sampling with a common setting of $\mathtt{top\_p}=1.0$ and $\mathtt{temperature}=1.0$.
As shown in \cref{tab:strategy}, using Nucleus Sampling slightly reduces the attack effectiveness.
However, even with this strategy, our method still forces the VLM to consume significantly more resources, producing much longer outputs than both the clean images and the other baseline attacks.

\subsection{Effect of Maximum Token Limit}
To test whether our attack effectiveness is bound by a specific output limit, we conducted an experiment comparing our default maximum generation length of 2048 tokens against a larger limit of 4096 tokens.
The results, presented in \cref{tab:4096}, clearly show that our attack potency scales directly with the VLM output capacity.
When the limit was increased to 4096, the average output length generated by our attack shows a significant increase.
These finding demonstrates that our method does not simply generate a fixed-length response.
Instead, it effectively forces the VLM into a non-terminating generative loop that is only halted by the external system constraint.

\section{Discussion}

\subsection{Transferability}
We provide the detailed results for transferability evaluation in \cref{tab:transferability}.
The target models include models from the same family but different scales (the 3B and 32B versions of Qwen2.5-VL) and models from entirely different developers (MiMo-VL, Gemma3-4B-IT, and Phi-Vision-Instruct).
The results consistently demonstrate that our attack exhibits little transferability.
We hypothesize that this stems from our attack's reliance on token-level optimization.
The internal representations and handling of special tokens are highly specific to each individual VLM, making the adversarial features non-transferable.
This inherent non-transferability makes the attack highly targeted, allowing an attacker to affect specific VLM services without causing unintended collateral impact on unrelated models.

%-------------------------------------------------------------------------------
\end{document}